\documentclass{optica-article}
\journal{opticajournal} % for journals or Optica Open

\articletype{Research Article}

\usepackage{lineno}
\begin{document}

\title{Complex-gate all-optical frequency-resolved optical gating for ultrabroadband isolated attosecond pulse characterization}

\author{Minshuang Xia,\authormark{1,2,3} Kaito Nishimiya,\authormark{1,4} Dianhong Dong,\authormark{1,2,3} Yuxi Fu, \authormark{2,3} and Eiji J. Takahashi\authormark{1,*}}

\address{
\authormark{1}Extreme Photonics Research Team, RIKEN Center for Advanced Photonics, RIKEN, 2-1, Hirosawa, Wako, Saitama 351-0198, Japan\\
\authormark{2}State Key Laboratory of Ultrafast Optical Science and Technology, Xi’an Institute of Optics and Precision Mechanics, Chinese Academy of Sciences, Xi’an 710119, China\\
\authormark{3}University of Chinese Academy of Sciences, Beijing 100049, China\\
}

\email{\authormark{4}kaito.nishimiya@riken.jp}
\email{\authormark{*}ejtak@riken.jp}

\begin{abstract*}
We experimentally demonstrate all-optical frequency-resolved optical gating (AO-FROG) for the characterization of ultrabroadband isolated attosecond pulses (IAPs) generated by a mid-infrared sub-cycle laser field. By extending the AO-FROG framework beyond the conventional phase-only modulation approximation, we develop a strong-field approximation (SFA)-based theoretical framework and show that the weak perturbing field induces both phase and amplitude modulations during high-order harmonic generation. A complex-valued gate function is therefore required for accurate pulse reconstruction.
Using a 2.26-$\mu$m sub-cycle driving laser, we characterize IAPs spanning 100--180 eV in argon. The measured AO-FROG traces exhibit delay-dependent spectral modulations arising from perturbation-induced modifications of the electron trajectories and ionization probability. SFA simulations reproduce the experimental observations and confirm the importance of including ionization-induced amplitude modulation. The reconstructed temporal and spectral properties reveal an IAP duration of approximately 300 as and its spectral phase, providing access to the attosecond chirp of the generated pulses.
Our results establish AO-FROG as a promising approach for temporal characterization of ultrabroadband attosecond sources driven by long-wavelength infrared fields.
\end{abstract*}

\section{Introduction}
The generation of isolated attosecond pulses (IAPs) with broad spectral bandwidths has opened new opportunities for investigating and controlling ultrafast electronic dynamics \cite{Hentschel2001,Kienberger2004}. Few-cycle laser-driven high-order harmonic generation (HHG) provides a promising route toward intense, broadband attosecond radiation in the extreme-ultraviolet (XUV) and soft X-ray spectral regions. However, accurate characterization of such pulses remains challenging because their ultrashort duration and broad bandwidth impose stringent requirements on the reconstruction methods \cite{Kim2014}.

Frequency-resolved optical gating for complete reconstruction of attosecond bursts (FROG-CRAB) \cite{Mairesse2005}, based on photoelectron measurements, has been widely used for attosecond pulse characterization. However, its reliance on electron detection and the central momentum approximation limits its applicability, particularly for broadband attosecond sources.
Recently, a collinear \textit{in situ} measurement technique, termed all-optical FROG (AO-FROG) \cite{Yang2020}, has been proposed as an alternative approach.
In AO-FROG, a weak perturbing laser field is introduced to modulate the HHG process, thereby encoding temporal information about the attosecond emission into the harmonic spectrum. By relying solely on optical detection, AO-FROG eliminates the need for photoelectron momentum measurements and offers a potentially broadband approach for characterizing attosecond pulses.

The HHG process involves three essential steps: tunnel ionization, electron propagation in the laser field, and recombination with the parent ion.
In previous collinear \cite{Dudovich2006,Doumy2009} and non-collinear \cite{Kim2013,Kim2013Nat.Photon.,Li2018,He2022} \textit{in situ} measurements, as well as in AO-FROG studies \cite{Yang2020,Meng2023,Dong2025,Meng2026}, the perturbing field has generally been treated as a pure-phase gate. Under this treatment, the primary effect of the weak perturbing field is attributed to the modification of the electron trajectory during the continuum propagation step of HHG. Within this approximation, the delay-dependent spectral modulation is interpreted as a temporal phase modulation of the attosecond emission. Although this model has successfully described several AO-FROG measurements \cite{Yang2020,Dong2025,Meng2026}, its validity for ultrabroadband attosecond pulses generated by few-cycle drivers remains unclear. Moreover, the gate function itself has received relatively little attention, and its role and significance in describing the spectral modulation have not been systematically investigated or evaluated.

Recent studies of perturbative high-harmonic wave mixing \cite{Wei2024} and \textit{in situ} measurements \cite{Yan2025} have provided evidence of amplitude modulation in addition to phase modulation, indicating that the perturbing field can also modify the instantaneous ionization rate through the strong-field tunneling process.
However, these observations have not yet been incorporated into a rigorous FROG-formalism description, nor has the corresponding gate been quantitatively derived and characterized. Consequently, the gate employed in AO-FROG retrieval lacks a clear physical description, limiting a comprehensive understanding of the underlying modulation mechanisms and the reliable characterization of ultrabroadband attosecond pulses. A rigorous description of the AO-FROG gate, incorporating both amplitude and phase contributions, is therefore essential for establishing the applicability and reliability of AO-FROG in the ultrabroadband regime.

In this work, we establish a rigorous theoretical framework for AO-FROG by deriving the dipole-moment response. The derivation reveals that the perturbing field introduces a complex gate function comprising both amplitude and phase modulation. The amplitude modulation originates from perturbation-induced variations in the strong-field ionization probability, whereas the phase modulation results from modifications to the electron propagation action. This generalized description provides a more complete picture of the AO-FROG process beyond the conventional phase-only approximation. Furthermore, we analyze the individual effects of amplitude and phase modulation on the AO-FROG trace and demonstrate how the amplitude and phase of the gate can be directly extracted from the measured delay-dependent spectra.

Based on this framework, we experimentally demonstrate AO-FROG characterization of IAPs generated by an ultrashort sub-cycle mid-infrared (MIR) driving field. The reconstructed attosecond pulses have a duration of approximately 300 as, and their spectral phase is retrieved, enabling evaluation of the attosecond chirp (atto-chirp).
Our results extend AO-FROG to MIR-driven broadband attosecond sources and establish a comprehensive approach for understanding and retrieving the perturbing-field-induced modulation in HHG. The developed complex-gate AO-FROG framework provides a general route toward accurate temporal characterization of next-generation attosecond light fields.

\section{Theoretical analysis}
\subsection{AO-FROG theory based on SFA}
In conventional FROG measurements, the gate function is determined by the nonlinear interaction between the pulse and the gating field. In AO-FROG, however, the gate is not an external optical response function but originates from the modification of the microscopic HHG emission process by the weak perturbing field.
Therefore, a theoretical description of the AO-FROG gate requires connecting the perturbation-induced modulation to the underlying electron dynamics during HHG.
A weak perturbing field can influence all three steps of the three-step model, leading to modifications in both the amplitude and phase of the emitted harmonic dipole.

Within the strong-field approximation (SFA) framework \cite{Lewenstein1994,Le2016}, we further apply the saddle-point approximation to the integrals over the canonical momentum $p$ and the ionization time $t^{\prime}$.
Solving
$p_{s}=\displaystyle \frac{1}{t-t^{\prime}_s }{\displaystyle \int }_{{t}^{\prime }_s}^{t}A(t^{\prime\prime} ){\mathrm{d}}{t}^{\prime\prime}$ yields the factor
$\left( \dfrac{-2\pi i}{t-t^\prime_s} \right)^{3/2}$,
which describes the quantum-diffusion effect \cite{Le2016}.
Because the solution of $\frac{1}{2}{[p_{s}-A(t^{\prime}_s )]}^{2}=-{I}_{p}$ leads to a complex ionization time, its real part satisfies $\frac{1}{2}{[p_{s}-{A}(t^{\prime}_s )]}^{2}=0$ \cite{Ivanov1996}.
Integrating over the real part of the ionization time for the quasi-classical action $S(p,t,t^{\prime})\approx S(p,t,t_s^{\prime})-I_p(t^{\prime}-t_s^{\prime})-\frac{1}{6}[E(t_s^{\prime})]^2(t^{\prime}-t_s^{\prime})^3$ yields another factor \cite{Wei2024},
$\sqrt{\pi}\left( \frac{2}{I_p |E(t_s^\prime)|^2} \right)^{\frac{1}{4}}
\exp{\left(-\frac{\left(2I_p\right)^\frac{3}{2}}{3\left|E\left(t_s^\prime\right)\right|}\right)},$
which describes the ionization contribution and has the same exponential dependence as that of the ADK model \cite{Ammosov1986}.
The unperturbed dipole moment along the $x$-axis can then be expressed as
\begin{equation}    
D_0\left(t\right)\propto\frac{d_x^\ast\left(p_s-A\left(t\right)\right)d_x\left(p_s-A\left(t_s^\prime\right)\right)E\left(t_s^\prime\right)}{\left|t-t_s^\prime\right|^\frac{3}{2} \sqrt{|E(t_s^\prime)|}}\exp{\left(-\frac{\left(2I_p\right)^\frac{3}{2}}{3\left|E\left(t_s^\prime\right)\right|}\right)}\exp{\left[-iS_0\left(p_s,t,t_s^\prime\right)\right]}+c.c.
\end{equation}
where the unperturbed quasi-classical action is
\begin{equation}
    S_0\left(p_s,t,t_s^\prime\right)={\int}_{t_s^\prime}^tdt^{\prime\prime}\left(\frac{1}{2}\left[p_s-A\left(t^{\prime\prime}\right)\right]^2+I_p\right).
\end{equation}

We now introduce a perturbing laser pulse $E_p(t)$ polarized along the $x$-axis, defined as $E_p(t)=\varepsilon E(t)$, where the electric-field ratio satisfies $\varepsilon \ll 1$.
The total electric field and vector potential are then given by $E_{\operatorname{total}}(t,\tau)= E(t)+E_p(t-\tau)$ and $A_{\operatorname{total}}(t,\tau)= A(t)+A_p(t-\tau)$, respectively, where $\tau$ is the relative delay between the fundamental driving pulse and the perturbing pulse.

We assume that the saddle-point solutions $p_s$ and $t_s^{\prime}$ remain unchanged in the presence of the perturbing pulse. The dipole moment in the presence of a delayed perturbing pulse $E_p(t-\tau)$ can then be written as
\begin{equation}    
D\left(t,\tau\right)\propto\frac{d_x^\ast\left(p_s-A\left(t\right)\right)d_x\left(p_s-A\left(t_s^\prime\right)\right)E\left(t_s^\prime\right)}{\left|t-t_s^\prime\right|^\frac{3}{2} \sqrt{|E(t_s^\prime)|}}\exp{\left(-\frac{\left(2I_p\right)^\frac{3}{2}}{3\left|E\left(t_s^\prime\right)+E_p(t_s^\prime-\tau)\right|}\right)}\exp{\left[-iS\left(p_s,t,t_s^\prime,\tau\right)\right]}+c.c.
\end{equation}
Using the first-order expansion $f(E+\Delta E)\approx f(E)+f^{\prime}(E) \cdot \Delta E$ and neglecting the second-order term in $A_p(t-\tau)$, we obtain \cite{Wei2024}
\begin{equation} \label{Eq: perturbing dipole}
D\left(t,\tau\right)=D_0\left(t\right)
\left[ 1+\frac{\left(2I_p\right)^\frac{3}{2}}{3\left|E\left(t_s^\prime\right)\right| E\left(t_s^\prime\right)} E_p(t_s^\prime-\tau)
\right]
\exp{ \left[i{\int}_{t_s^\prime}^tdt^{\prime\prime}\left[p_s-A\left(t^{\prime\prime}\right)\right]A_p(t^{\prime\prime}-\tau)\right]}
+c.c.
\end{equation}
This expression indicates that the effective gate is complex and can be described in the form given in Ref.~\cite{Kim2013} as
\begin{equation}
 G(t,\tau)=[1+\alpha(t,\tau)]\exp{\left[i\sigma(t,\tau)\right]},
\end{equation}
where $\alpha(t,\tau)$ represents the perturbing-pulse-induced amplitude modulation and $\sigma(t,\tau)$ represents the additional phase induced by the perturbing pulse.
Assuming that the excursion time $\Delta t=t-t_s^\prime$ remains unchanged, we obtain
\begin{eqnarray}
\alpha(t,\tau)=\frac{\left(2I_p\right)^\frac{3}{2}}{3\left|E\left(t_s^\prime\right)\right| E\left(t_s^\prime\right)} E_p(t_s^\prime-\tau)
=\frac{\left(2I_p\right)^\frac{3}{2}}{3\left|E\left(t_s^\prime\right)\right| E\left(t_s^\prime\right)} E_p(t-\tau-\Delta t)=\alpha(t-\tau),
\end{eqnarray}
\begin{eqnarray}
\sigma(t,\tau) &=&{\int}_{t_s^\prime}^tdt^{\prime\prime}\left[p_s-A\left(t^{\prime\prime}\right)\right]A_p(t^{\prime\prime}-\tau) \approx {\int}_{t-\Delta t}^t dt^{\prime\prime} p_s A_p(t^{\prime\prime}-\tau) \notag \\
&=& p_s{\int}_{t-\tau-\Delta t}^{t-\tau} dt^{\prime\prime}  A_p(t^{\prime\prime}) = \sigma(t-\tau),
\end{eqnarray}
where we neglect the $2\omega$ oscillatory contribution arising from the product $A(t^{\prime\prime})A_p(t^{\prime\prime}-\tau)$.
The AO-FROG trace can therefore be expressed in the standard FROG form as 
\begin{eqnarray}\label{Eq: AO-FROG trace}
I_{\mathrm{AO-FROG}} (\omega , \tau)
&=&\omega^4 \left| \int_ {- \infty} ^ {\infty} D(t,\tau) \exp (- i \omega t) d t \right| ^ {2} 
=\omega^4 \left| \int_ {- \infty} ^ {\infty} D_0 (t) G (t - \tau) \exp (- i \omega t) d t \right| ^ {2}  \notag\\
&=&\omega^4 \left| \int_ {- \infty} ^ {\infty} D_0 (t) [1+\alpha(t-\tau)]\exp{\left[i\sigma(t-\tau)\right]} \exp (- i \omega t) d t \right| ^ {2}.
\end{eqnarray}

Previous descriptions of AO-FROG have primarily considered phase modulation arising from the perturbation of continuum-electron propagation. Here, based on our theoretical derivation, we demonstrate that perturbation-induced modulation of the ionization probability provides an additional contribution that is essential for accurately describing the AO-FROG trace.
Therefore, we further analyze the respective roles of phase and amplitude modulation and investigate how these two mechanisms jointly influence the measured AO-FROG trace.

\subsection{Roles of amplitude and phase modulations in AO-FROG}

FROG enables the complete characterization of ultrashort optical fields by retrieving their temporal amplitude and phase from a delay-dependent time-frequency spectrogram \cite{Trebino1997}. The general FROG trace is expressed as
\begin{equation}
I_{\mathrm{FROG}}(\omega,\tau)=
\left|
\int_{-\infty}^{\infty}
E(t)G(t-\tau)e^{-i\omega t}dt
\right|^2
    =\left| \int_ {- \infty} ^ {\infty} E_{\mathrm{sig}} (t,\tau) e^{-i\omega t} d t \right| ^ {2},
\end{equation}
where $E(t)$ is the unknown pulse and $G(t-\tau)$ is a delay-dependent temporal gate. The gate plays a central role in mapping temporal information onto measurable spectral modulation and can introduce amplitude modulation, phase modulation, or both.

\begin{figure}[htbp]
\centering     
\includegraphics[width=1\linewidth]{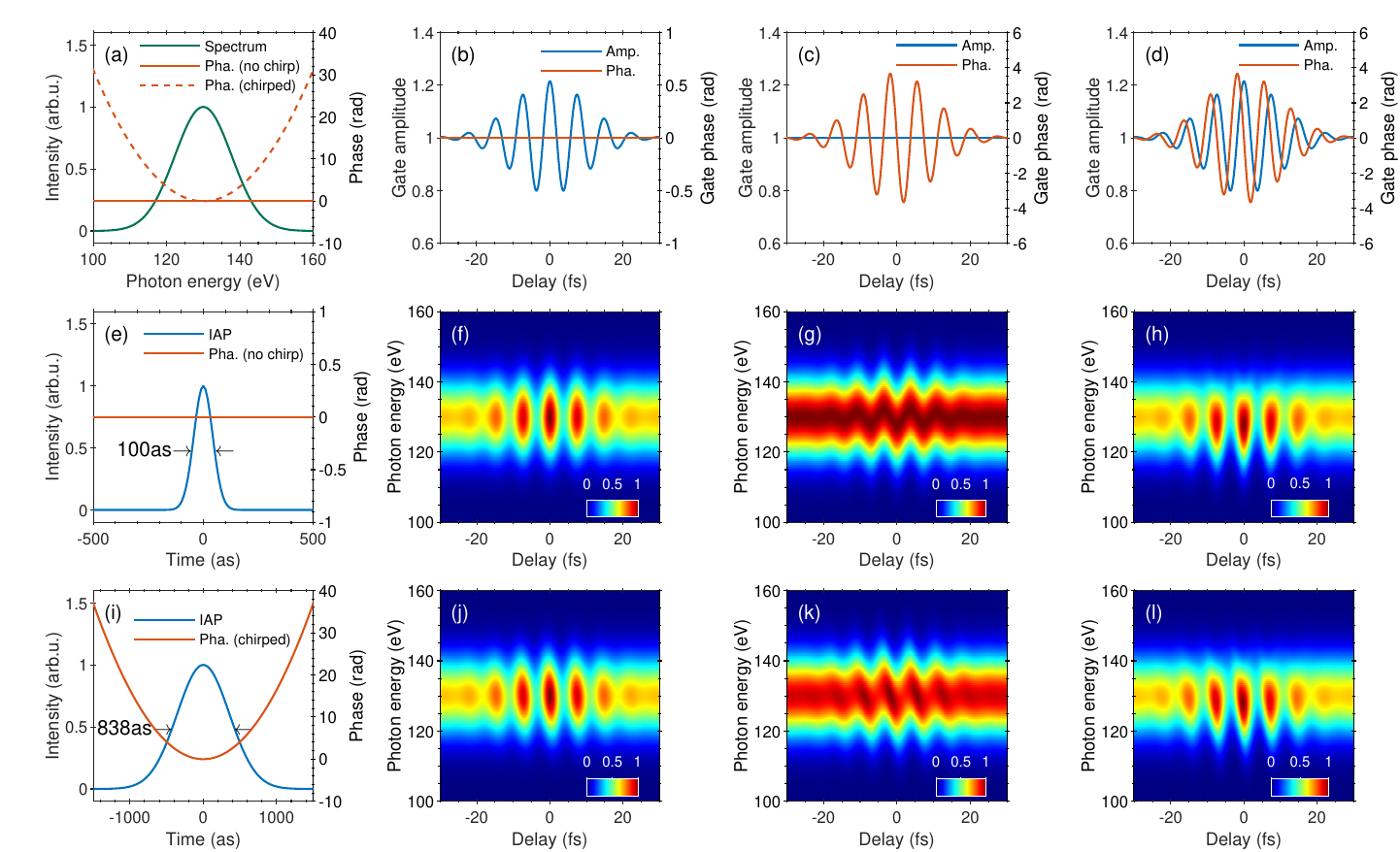}     
\caption{Simulated FROG-type traces obtained using different gates by LSGPA. (a) Spectrum and spectral phase of unchirped and chirped ($3\times10^4~\mathrm{as}^2$) Gaussian IAPs. (b)--(d) Pure-amplitude, pure-phase, and complex gates, respectively. (e) Temporal profile and temporal phase of the unchirped IAP. (f)--(h) Simulated traces of the unchirped IAP using pure-amplitude, pure-phase, and complex gates, respectively. (i) Temporal profile and temporal phase of the chirped IAP. (j)--(l) Simulated traces of the chirped IAP using pure-amplitude, pure-phase, and complex gates, respectively.} 
\label{Fig: different gates} 
\end{figure}

The modulation mechanism depends on the underlying nonlinear process. In conventional second-harmonic-generation FROG, the nonlinear interaction produces a signal field proportional to the pulse intensity, resulting primarily in amplitude modulation of the spectrogram. The FROG traces of the unchirped and chirped IAPs with a pure-amplitude gate are shown in Figs.~\ref{Fig: different gates}(f) and \ref{Fig: different gates}(j), respectively. For attosecond pulse characterization, conventional nonlinear optical gating is generally not applicable because attosecond pulses are typically generated in the XUV region. Instead, FROG-CRAB employs a synchronized streaking field that primarily modifies the photoelectron momentum while having a comparatively weak effect on the ionization probability, corresponding to a predominantly phase gate, as shown in Figs.~\ref{Fig: different gates}(g) and \ref{Fig: different gates}(k) \cite{Mairesse2005,Kim2014}.

In AO-FROG, the perturbing laser field acts during the HHG process, and the emitted harmonic dipole moment can be expressed as
$D(t,\tau)=D_0(t)G(t,\tau)$.
Unlike FROG-CRAB, the perturbing field influences both the electron trajectories and the ionization probability. Consequently, the gate $G(t)=\left[1+\alpha(t)\right]e^{i\sigma(t)}$ is generally complex, where $\alpha(t)$ and $\sigma(t)$ denote the amplitude and phase modulations, respectively.

To illustrate the roles of the two modulation mechanisms, Fig.~\ref{Fig: different gates} provides a general illustration of the underlying modulation effects and is independent of the specific experimental conditions discussed below. Figs.~\ref{Fig: different gates}(f--h) and \ref{Fig: different gates}(j--l) compare simulated FROG traces generated using pure-amplitude, pure-phase, and complex gates. Amplitude modulation primarily redistributes the harmonic intensity, whereas phase modulation produces delay-dependent spectral shifts. Only their combination reproduces the characteristic modulation pattern observed in AO-FROG traces, demonstrating that both amplitude and phase modulation are required for reliable characterization of ultrabroadband attosecond pulses.

\subsection{Simulation and reconstruction results}

To verify the validity of AO-FROG for the characterization of ultrabroadband IAPs, we perform numerical simulations for argon using the SFA. The simulated AO-FROG trace is generated using a 1-cycle, 2.26-$\mu$m driving pulse with a CEP of 0 and an intensity of $1.5\times10^{14}\,\mathrm{W/cm^2}$. Here, the 1-cycle pulse is adopted as a simplified approximation to the 6.7-fs sub-cycle pulse used in the experiment.
The perturbing pulse has the same temporal profile as the driving pulse but a lower intensity of $7.5\times10^{11}\,\mathrm{W/cm^2}$. The corresponding electric-field ratio is $\varepsilon \approx 0.071$, which is sufficiently small for the first-order expansion used in Eq.~\eqref{Eq: perturbing dipole} to remain a valid approximation.

Inspired by femtosecond FROG \cite{DeLong1994,Trebino2000}, we first calculate the delay marginal as
$M(\tau)=\int_{-\infty}^{\infty} I(\omega,\tau)\,\mathrm{d}\omega.$
The delay-dependent central photon energy is calculated as
\begin{equation}
E_{\mathrm{c}}(\tau)=
\frac{\displaystyle\int_{-\infty}^{\infty} \hbar\omega\,I(\omega,\tau)\,\mathrm{d}\omega}
{\displaystyle\int_{-\infty}^{\infty} I(\omega,\tau)\,\mathrm{d}\omega}.
\end{equation}
The delay marginal $M(\tau)$ primarily reflects the variation of the gate amplitude with delay and can therefore be used to estimate the gate amplitude.
The central photon energy $E_{\mathrm{c}}(\tau)$ reflects the delay-dependent shift of the spectral center of gravity and is more sensitive to the gate phase; it can therefore be used to analyze phase modulation.
Furthermore, the gate amplitude $\alpha(\tau)$ can be directly estimated from the relationship $M(\tau)\propto[1+\alpha(\tau)]^2$ \cite{DeLong1994}.
The shift in central photon energy with delay is related to the first-order variation of the gate phase. 
From the mathematical form in Eq.~\eqref{Eq: AO-FROG trace}, $\Delta E_{\mathrm{c}}(\tau)=\hbar\frac{\partial\sigma(\tau)}{\partial \tau}$ determines the shift in central photon energy \cite{DeLong1994}.
Therefore, the gate phase $\sigma(\tau)$ can be estimated by integrating the central photon-energy shift $\Delta E_{\mathrm{c}}(\tau)$.

\begin{figure}[htbp]
    \centering
    \includegraphics[width=0.8\linewidth]{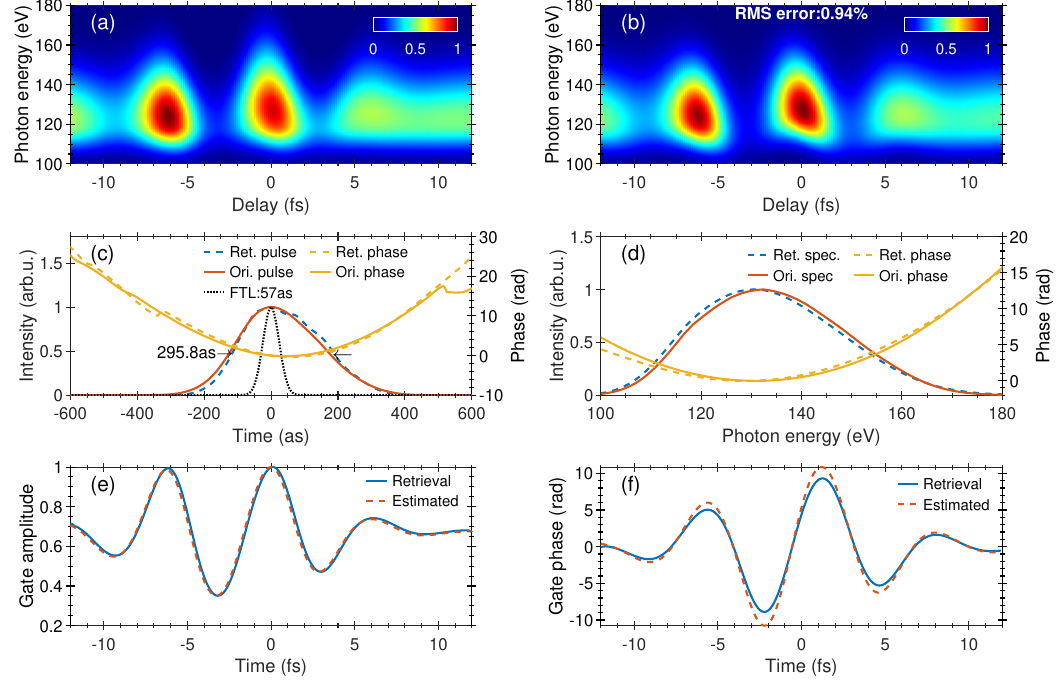}
    \caption{SFA simulation and reconstruction results for IAPs in argon. (a) Simulated AO-FROG trace. (b) Reconstructed AO-FROG trace with an RMS error of 0.94\%. 
    (c) Retrieved (blue dashed line), original (red solid line), and FTL (black dotted line) temporal profiles, together with the retrieved (yellow dashed line) and original (yellow solid line) temporal phases.
    (d) Retrieved (blue dashed line) and original (red solid line) spectra, together with the retrieved (yellow dashed line) and original (yellow solid line) spectral phases. A positive atto-chirp of 5200 as$^2$ is obtained by fitting the spectral phase.
    (e) Normalized reconstructed (blue solid line) and estimated (red dashed line) gate amplitudes. (f) Reconstructed (blue solid line) and estimated (red dashed line) gate phases.}
    \label{Fig: SFA AO-FROG}
\end{figure}

We select the experimentally relevant short-trajectory region by applying a window function to the dipole acceleration.
The simulated AO-FROG trace is shown in Fig.\ref{Fig: SFA AO-FROG}(a), obtained by scanning the relative delay between the driving and perturbing pulses.
The reference results in Figs.~\ref{Fig: SFA AO-FROG}(c) and \ref{Fig: SFA AO-FROG}(d) are calculated using the SFA without the perturbing pulse and show good consistency with the results retrieved from the simulated AO-FROG trace.
Moreover, the retrieved gate agrees well with the estimated gate. The relative delay between the gate amplitude and phase is approximately 0.36$\pi$. This relative delay can be interpreted as the time interval between electron ionization and recombination. Because the electron excursion time varies only slightly for electrons ionized at different times, the relative delay is expected to exhibit only weak variations.

\section{Experimental results and discussions}
\begin{figure}[htbp]
\centering     
\includegraphics[width=1\linewidth]{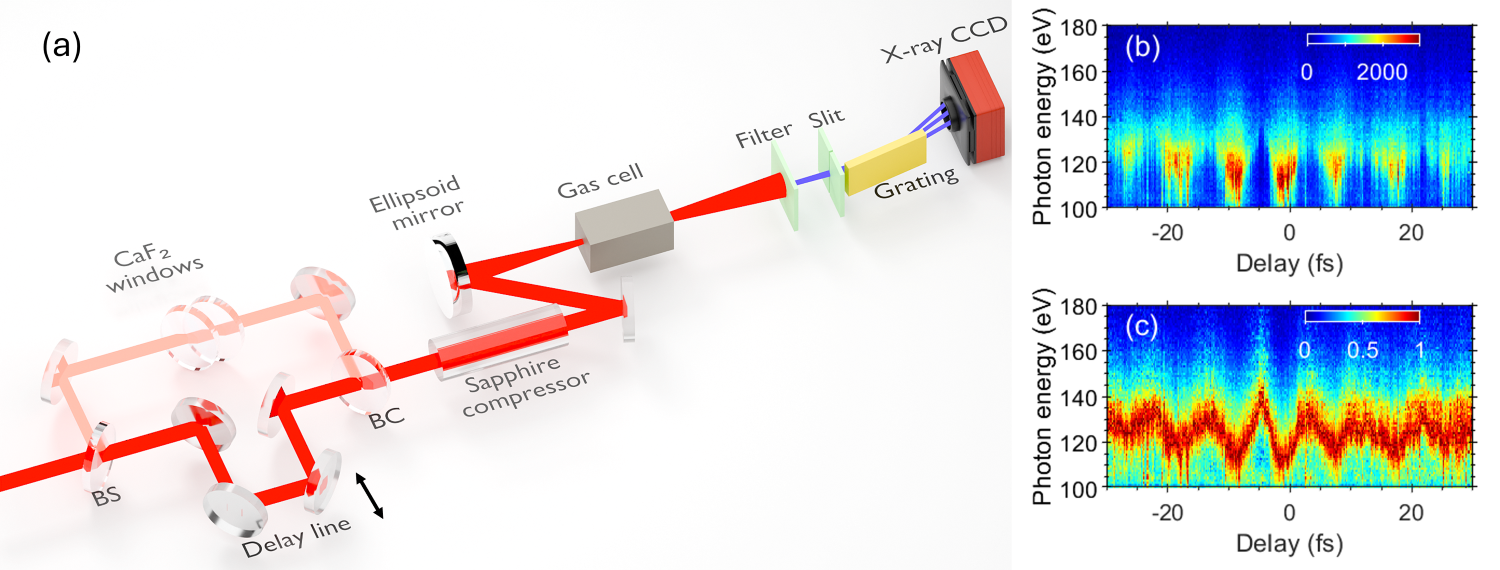}     
\caption{(a) Schematic of the experimental setup for AO-FROG measurement (BS, beam splitter; BC, beam combiner). The spectrometer consists of a slit, a flat-field grating, and an X-ray CCD. (b) Resampled spectrogram with a delay step of 200 as. (c) Normalized AO-FROG trace.} 
\label{Fig: experimental setup} 
\end{figure}

A MIR sub-cycle laser system (50 mJ, 6.7 fs, 2.26 $\mu$m, 10 Hz), based on a synthesized seed generator and heterogeneous nonlinear-crystal dual-chirped optical parametric amplification (DC-OPA), is employed as the driving source \cite{Nishimiya2025,Nishimiya2025_arXiv}. As shown in Fig.~\ref{Fig: experimental setup}(a), the experimental setup is based on a Mach--Zehnder interferometer. The S-polarized driving pulse is split and recombined using two single-side anti-reflection-coated CaF$_2$ windows, producing a weak perturbing pulse with an intensity ratio of 1:0.005. Two CaF$_2$ windows are inserted into the perturbing arm to compensate for the dispersion mismatch between the two interferometer arms. The relative delay is adjusted using a piezo-driven translation stage in the main-pulse arm. The combined beam is compressed using a sapphire bulk compressor and then focused into a 10-mm argon gas cell using an ellipsoidal mirror with a focal length of 5 m.
Figure~\ref{Fig: experimental setup}(b) shows the measured delay-dependent HHG spectra in argon at a driving intensity of $1.5\times10^{14}\,\mathrm{W/cm^2}$ and a perturbing intensity of $7.5\times10^{11}\,\mathrm{W/cm^2}$, corresponding to an electric-field ratio of $\varepsilon \approx 0.071$.
The measured spectrogram exhibits pronounced spectral modulation, indicating that the attosecond emission is encoded by the weak perturbing field.
To reveal the underlying phase modulation, the spectrum at each delay is normalized, as shown in Fig.~\ref{Fig: experimental setup}(c). After normalization, a clear delay-dependent shift of the spectral centroid becomes apparent, which is largely obscured by the intensity variation in the original trace.
The simultaneous observation of intensity modulation and spectral centroid shifts provides experimental evidence that the perturbing field modifies both the harmonic amplitude and spectral phase. Therefore, the experimental AO-FROG trace provides direct evidence that the gate is inherently complex-valued rather than purely phase-only, resulting in simultaneous amplitude and phase modulation during the gating process.

Using the least-squares generalized projection algorithm (LSGPA) \cite{Gagnon2008}, we reconstruct the temporal structure of the IAP with a low root-mean-square (RMS) error.
The RMS error between the retrieved and measured AO-FROG traces is defined as ${\mathrm{RMSE}}=\sqrt{\frac{1}{N}\sum|I_{\mathrm{ret}}-I_{\mathrm{mea}}|^2}=\sqrt{\frac{1}{N}\sum|I_{\mathrm{ret}}-I_{\mathrm{true}}-I_{\mathrm{noise}}|^2}$. In LSGPA, the least-squares procedure approximately minimizes the RMS error at each iteration \cite{Gagnon2008}.

Although the RMS error generally decreases during iterative optimization, a lower RMS error does not necessarily indicate a more accurate reconstruction. After a sufficient number of iterations, the optimization may begin to fit experimental noise rather than the underlying signal, resulting in a further reduction of the RMS error while degrading the physical reliability of the retrieved pulse \cite{Trebino2000,Hansen2010}.
Therefore, the stopping criterion should balance error reduction with the physical reliability of the reconstruction \cite{Kane1999}. In this work, the maximum number of iterations is set to 2000. The delay range is from $-30$ to $30$ fs with a step size of 200 as, consistent with the experimental measurements, while the temporal range of the retrieved pulse is from $-2000$ to $2000$ as with a temporal step size of 8 as.
The retrieval is terminated when the relative change in the RMS error remains below $10^{-3}$ for 100 consecutive iterations, indicating that the least-squares residual has essentially reached the experimental noise floor \cite{Kane1999}.

\begin{figure}[htbp]
    \centering
    \includegraphics[width=0.8\linewidth]{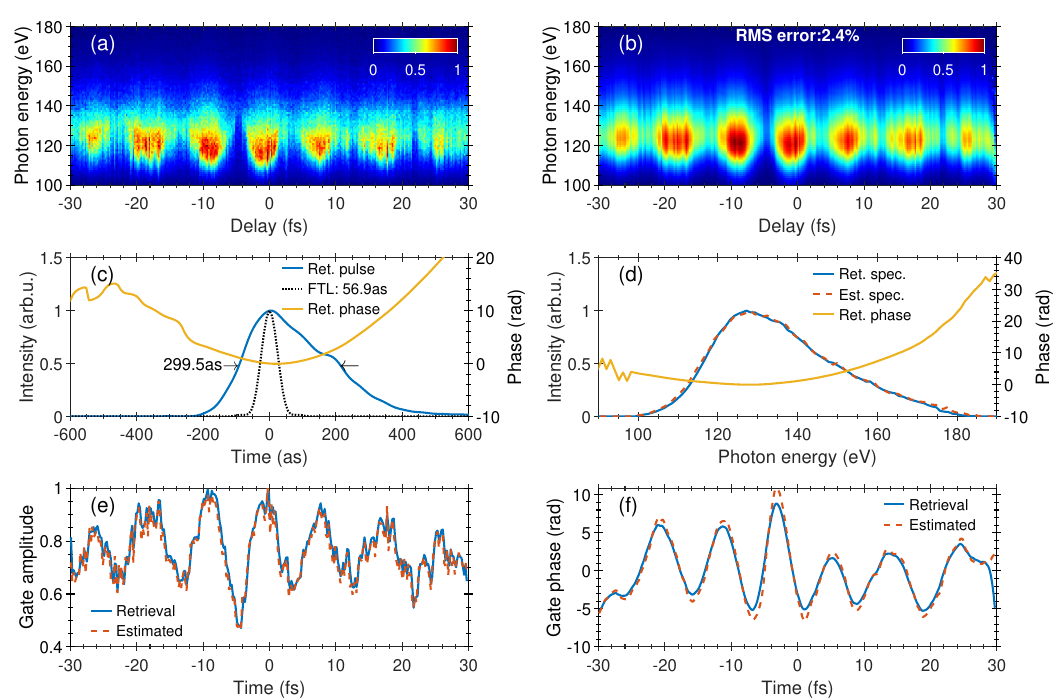}
    \caption{Temporal characterization of the generated IAPs in argon using AO-FROG. (a) Measured two-dimensional AO-FROG trace. (b) Reconstructed AO-FROG trace with an RMS error of 2.4\%. (c) Retrieved temporal profile (blue solid line), Fourier-transform-limited pulse profile (black dotted line), and temporal phase (yellow solid line). (d) Retrieved spectrum (blue solid line), estimated spectrum (red dashed line), and retrieved spectral phase (yellow solid line). A positive atto-chirp of 5800 as$^2$ is obtained by fitting the spectral phase. (e) Normalized reconstructed (blue solid line) and estimated (red dashed line) gate amplitudes. (f) Reconstructed (blue solid line) and estimated (red dashed line) gate phases.}
    \label{Fig: exp AO-FROG}
\end{figure}

As shown in Fig.~\ref{Fig: exp AO-FROG}(a), the experimentally measured AO-FROG trace is obtained after numerically applying a 100-nm Ag filter and is used as the input for IAP retrieval. The spectral range of 100--180 eV is therefore defined by the experimentally accessible spectral region after filtering. The retrieved AO-FROG trace, reconstructed temporal and spectral profiles of the IAP, and complex gate are presented in Fig.~\ref{Fig: exp AO-FROG}.
The reconstructed trace shown in Fig.~\ref{Fig: exp AO-FROG}(b) reproduces the experimental measurement well, with an RMS error of 2.4\%, indicating good convergence of the retrieval algorithm.
The retrieved results shown in Figs.~\ref{Fig: SFA AO-FROG} and \ref{Fig: exp AO-FROG} consistently demonstrate that amplitude and phase modulations coexist in AO-FROG and jointly give rise to the distinct modulation features observed in the measured trace. Moreover, the retrieved gate exhibits a complex-valued form, in agreement with the theoretical prediction. Its amplitude and phase components correspond to the two distinct modulation mechanisms encoded in the AO-FROG trace, providing experimental evidence for the complex nature of the gate. The oscillation periods of both the estimated gate amplitude and phase are approximately 8.3 fs.

The retrieved temporal profile yields an IAP with a full width at half-maximum duration of 299.5 as, as shown in Fig.~\ref{Fig: exp AO-FROG}(c). The retrieved spectral intensity is presented in Fig.~\ref{Fig: exp AO-FROG}(d). For comparison, the spectral profile estimated directly from the measured AO-FROG trace is obtained from the frequency marginal $M(\omega)=\int_{-\infty}^{\infty} I(\omega,\tau)\,\mathrm{d}\tau$ and plotted as the red dashed curve. The frequency marginal provides an independent estimate of the spectral intensity without relying on the retrieval process. The good agreement between the retrieved spectrum and the frequency marginal further supports the consistency and reliability of the reconstruction.

The reconstruction results obtained from the experimental measurements and SFA simulations show good agreement. Specifically, the normalized gate amplitudes in Figs.~\ref{Fig: exp AO-FROG}(e) and \ref{Fig: SFA AO-FROG}(e) both reach minimum values of approximately 0.4, while the reconstructed gate phases in Figs.~\ref{Fig: exp AO-FROG}(f) and \ref{Fig: SFA AO-FROG}(f) span a similar range of approximately $-8$ to 8 rad. These results further demonstrate that the SFA model captures the essential features observed in the experimental reconstruction.

Also, we estimate the atto-chirp from both the reconstructed spectral phase and the scaling law.
The atto-chirp of high harmonics generated by the short electron trajectory is positive, and its value near the center of the plateau region can be approximately estimated using the scaling law \cite{Chang2019}:
\begin{equation} \label{Eq: chirp}
    \mathrm{Chirp}[\mathrm{as^2}]= ({1.63\times 10^{18}})/({I_0[\mathrm{W/cm^2}]\cdot \lambda_0[\mu \mathrm{m}]}),
\end{equation}
where $I_0$ and $\lambda_0$ are the driving laser intensity and wavelength, respectively.
The validity of this scaling law can be assessed using the experimental results reported in Ref.~\cite{Feng2009}. For a driving pulse with an intensity of $2\times10^{14}$ $\mathrm{W/cm^2}$ and a wavelength of 790 nm, Eq.~\eqref{Eq: chirp} predicts an atto-chirp of $1.03\times10^4$ as$^2$, which is reasonably close to the value of 7850 as$^2$ retrieved using FROG-CRAB. The slight discrepancy is expected because the atto-chirp decreases toward the cutoff region, whereas 
Eq.~\eqref{Eq: chirp} provides an approximate estimate for the plateau region.

Applying Eq.~\eqref{Eq: chirp} to the experimental parameters used in this work ($I_0=1.5\times10^{14}~ \mathrm{W/cm^2}$ and $\lambda_0=2.26~\mu \mathrm{m}$) yields an estimated atto-chirp of 4850 as$^2$.
Independently, the reconstructed spectral phase is fitted with a quadratic function $\varphi(\omega)=a+b(\omega-\omega_0)+c(\omega-\omega_0)^2$, and the corresponding atto-chirp is obtained from the second-order coefficient as approximately 5800 as$^2$. The retrieved value is in reasonable agreement with the SFA prediction of 5200 as$^2$, indicating that the reconstructed spectral phase is consistent with the expected atto-chirp of the generated IAP.
At the same time, the retrieved atto-chirp of $5800~\mathrm{as}^2$, corresponding to $8.8~\mathrm{as/eV}$ ($1~\mathrm{as/eV}=658.2~\mathrm{as}^2$), is too small to produce a visually distinguishable slope in the AO-FROG trace. Even across a spectral range of $100~\mathrm{eV}$, the corresponding temporal shift is only $880~\mathrm{as}$, equivalent to approximately four delay pixels at the delay step size of $200~\mathrm{as}$. Moreover, the coexistence of amplitude and phase modulation further masks the weak signature of the atto-chirp, explaining why no obvious slope can be identified directly from the experimental AO-FROG trace.

\section{Conclusion and outlook}
In conclusion, we have developed an SFA-based theoretical framework for AO-FROG, demonstrating that the AO-FROG gate is intrinsically complex-valued and comprises both amplitude and phase components, which jointly determine the measured AO-FROG trace. We further elucidate the respective effects of amplitude and phase modulation on the AO-FROG trace and demonstrate that the amplitude and phase of the complex gate can be directly extracted from the marginals of the measured trace. By combining experimental measurements with SFA simulations, we demonstrate the feasibility and reliability of AO-FROG for characterizing ultrabroadband IAPs generated by a MIR sub-cycle driving laser.

To the best of our knowledge, this work provides the first experimental demonstration of AO-FROG characterization of supercontinuum IAPs driven by a MIR sub-cycle laser, as well as the first direct experimental and theoretical reconstruction of the complex gate in AO-FROG. These findings establish the potential of AO-FROG for characterizing attosecond sources driven by long-wavelength laser fields and provide a promising route toward more physically accurate characterization of ultrabroadband attosecond pulses.

Looking forward, the rapid development of high-energy DC-OPA systems is expected to enable the generation of IAPs extending toward the water-window regime. The complex-gate AO-FROG framework presented here therefore provides a promising all-optical temporal metrology approach for next-generation ultrabroadband, high-photon-energy attosecond light sources.
In principle, the complex-gate AO-FROG approach is not intrinsically restricted by the photon energy of the generated attosecond pulses. Nevertheless, extending this technique to higher photon energies requires careful consideration of the approximations adopted in the theoretical description, such as the assumption that the electron excursion time remains unchanged during the propagation process. Notably, this approximation is expected to become increasingly valid in the higher-energy regime. These developments could establish AO-FROG as a versatile temporal metrology tool for ultrabroadband attosecond sources beyond the current regime.

\begin{backmatter}
\bmsection{Funding}
We acknowledge financial support from the Ministry of Education, Culture, Sports, Science and Technology of Japan (MEXT) Quantum Leap Flagship Program (Q-LEAP) (grant no. JP-MXS0118068681), the Special Postdoctoral Researchers Program of RIKEN, and the RIKEN Incentive Research Project.

\bmsection{Acknowledgment}
M. X. and D. D. thank the International Program Associate Program of RIKEN for support.

\bmsection{Disclosures}
The authors declare no conflicts of interest.

\bmsection{Data availability} 
Data underlying the results presented in this paper are not publicly available at this time but may be obtained from the authors upon reasonable request.

\end{backmatter}

\bibliography{reference}

@article{Hentschel2001,
   author = {Hentschel, M. and Kienberger, R. and Spielmann, C. and Reider, G. A. and Milosevic, N. and Brabec, T. and Corkum, P. and Heinzmann, U. and Drescher, M. and Krausz, F.},
   title = {Attosecond metrology},
   journal = {Nature},
   volume = {414},
   number = {6863},
   pages = {509-513},
   ISSN = {0028-0836},
   DOI = {Doi 10.1038/35107000},
   url = {<Go to ISI>://WOS:000172405900038
https://www.nature.com/articles/35107000.pdf},
   year = {2001},
   type = {Journal Article}
}

@article{Kienberger2004,
   author = {Kienberger, R. and Goulielmakis, E. and Uiberacker, M. and Baltuska, A. and Yakovlev, V. and Bammer, F. and Scrinzi, A. and Westerwalbesloh, T. and Kleineberg, U. and Heinzmann, U. and Drescher, M. and Krausz, F.},
   title = {Atomic transient recorder},
   journal = {Nature},
   volume = {427},
   number = {6977},
   pages = {817-821},
   ISSN = {0028-0836},
   DOI = {10.1038/nature02277},
   url = {<Go to ISI>://WOS:000189207500033
https://www.nature.com/articles/nature02277.pdf},
   year = {2004},
   type = {Journal Article}
}

@article{Doumy2009,
   author = {Doumy, G. and Wheeler, J. and Roedig, C. and Chirla, R. and Agostini, P. and DiMauro, L. F.},
   title = {Attosecond Synchronization of High-Order Harmonics from Midinfrared Drivers},
   journal = {Phys. Rev. Lett.},
   volume = {102},
   number = {9},
   ISSN = {0031-9007},
   DOI = {ARTN 093002
10.1103/PhysRevLett.102.093002},
   url = {<Go to ISI>://WOS:000263911900019
https://journals.aps.org/prl/pdf/10.1103/PhysRevLett.102.093002},
   year = {2009},
   type = {Journal Article}
}

@article{Kim2013Nat.Photon.,
   author = {Kim, K. T. and Zhang, C. M. and Shiner, A. D. and Schmidt, B. E. and Légaré, F. and Villeneuve, D. M. and Corkum, P. B.},
   title = {Petahertz optical oscilloscope},
   journal = {Nat Photonics},
   volume = {7},
   number = {12},
   pages = {958-962},
   ISSN = {1749-4885},
   DOI = {10.1038/nphoton.2013.347},
   url = {<Go to ISI>://WOS:000327738400012
https://www.nature.com/articles/nphoton.2013.347.pdf},
   year = {2013},
   type = {Journal Article}
}

@article{Chang2019,
   author = {Chang, Zenghu},
   title = {Compensating chirp of attosecond X-ray pulses by a neutral hydrogen gas},
   journal = {OSA Continuum},
   volume = {2},
   number = {2},
   pages = {314-319},
   ISSN = {2578-7519},
   DOI = {10.1364/osac.2.000314},
   year = {2019},
   type = {Journal Article}
}

@article{Dong2025,
   author = {Dong, Dianhong and Wang, Hushan and Xue, Bing and Imasaka, Kotaro and Kanda, Natsuki and Fu, Yuxi and Nabekawa, Yasuo and Takahashi, Eiji J.},
   title = {Perturbed three-channel waveform synthesizer for efficient isolated attosecond pulse generation and characterization},
   journal = {Opt. Lett.},
   volume = {50},
   number = {5},
   pages = {1461-1464},
   DOI = {10.1364/OL.551513},
   url = {https://opg.optica.org/ol/abstract.cfm?URI=ol-50-5-1461},
   year = {2025},
   type = {Journal Article}
}

@article{Dudovich2006,
   author = {Dudovich, N. and Smirnova, O. and Levesque, J. and Mairesse, Y. and Ivanov, M. Y. and Villeneuve, D. M. and Corkum, P. B.},
   title = {Measuring and controlling the birth of attosecond XUV pulses},
   journal = {Nat Phys},
   volume = {2},
   number = {11},
   pages = {781-786},
   ISSN = {1745-2473},
   DOI = {10.1038/nphys434},
   url = {<Go to ISI>://WOS:000242477600018
https://www.nature.com/articles/nphys434},
   year = {2006},
   type = {Journal Article}
}

@article{Feng2009,
   author = {Feng, X. M. and Gilbertson, S. and Mashiko, H. and Wang, H. and Khan, S. D. and Chini, M. and Wu, Y. and Zhao, K. and Chang, Z. H.},
   title = {Generation of Isolated Attosecond Pulses with 20 to 28 Femtosecond Lasers},
   journal = {Phys. Rev. Lett.},
   volume = {103},
   number = {18},
   ISSN = {0031-9007},
   DOI = {ARTN 183901
10.1103/PhysRevLett.103.183901},
   url = {<Go to ISI>://WOS:000271352400017
https://journals.aps.org/prl/pdf/10.1103/PhysRevLett.103.183901},
   year = {2009},
   type = {Journal Article}
}

@article{He2022,
   author = {He, L. X. and Hu, J. C. and Sun, S. Q. and He, Y. Q. and Deng, Y. and Lan, P. F. and Lu, P. X.},
   title = {All-optical spatio-temporal metrology for isolated attosecond pulses},
   journal = {J. Phys. B: At. Mol. Opt. Phys.},
   volume = {55},
   number = {20},
   ISSN = {0953-4075},
   DOI = {ARTN 205601
10.1088/1361-6455/ac8f01},
   url = {<Go to ISI>://WOS:000857980300001
https://iopscience.iop.org/article/10.1088/1361-6455/ac8f01/pdf},
   year = {2022},
   type = {Journal Article}
}

@article{Kim2014,
   author = {Kim, K. T. and Villeneuve, D. M. and Corkum, P. B.},
   title = {Manipulating quantum paths for novel attosecond measurement methods},
   journal = {Nat Photonics},
   volume = {8},
   number = {3},
   pages = {187-194},
   ISSN = {1749-4885},
   DOI = {10.1038/nphoton.2014.26},
   url = {<Go to ISI>://WOS:000332221100008
https://www.nature.com/articles/nphoton.2014.26.pdf},
   year = {2014},
   type = {Journal Article}
}

@article{Kim2013,
   author = {Kim, K. T. and Zhang, C. M. and Shiner, A. D. and Kirkwood, S. E. and Frumker, E. and Gariepy, G. and Naumov, A. and Villeneuve, D. M. and Corkum, P. B.},
   title = {Manipulation of quantum paths for space-time characterization of attosecond pulses},
   journal = {Nat Phys},
   volume = {9},
   number = {3},
   pages = {159-163},
   ISSN = {1745-2473},
   DOI = {DOI 10.1038/nphys2525},
   url = {<Go to ISI>://WOS:000316156300018
https://www.nature.com/articles/nphys2525.pdf},
   year = {2013},
   type = {Journal Article}
}

@article{Le2016,
   author = {Le, A. T. and Wei, H. and Jin, C. and Lin, C. D.},
   title = {Strong-field approximation and its extension for high-order harmonic generation with mid-infrared lasers},
   journal = {J. Phys. B: At. Mol. Opt. Phys.},
   volume = {49},
   number = {5},
   ISSN = {0953-4075},
   DOI = {Artn 053001
10.1088/0953-4075/49/5/053001},
   url = {<Go to ISI>://WOS:000370601400001
https://iopscience.iop.org/article/10.1088/0953-4075/49/5/053001/pdf},
   year = {2016},
   type = {Journal Article}
}

@article{Lewenstein1994,
   author = {Lewenstein, M. and Balcou, P. and Ivanov, M. Y. and L'Huillier, A. and Corkum, P. B.},
   title = {Theory of high-harmonic generation by low-frequency laser fields},
   journal = {Phys. Rev. A},
   volume = {49},
   number = {3},
   pages = {2117-2132},
   ISSN = {1050-2947 (Print)
1050-2947 (Linking)},
   DOI = {10.1103/physreva.49.2117},
   url = {https://www.ncbi.nlm.nih.gov/pubmed/9910464},
   year = {1994},
   type = {Journal Article}
}

@article{Li2018,
   author = {Li, Z. Y. and Kong, F. Q. and Brown, G. and Hammond, T. J. and Ko, D. H. and Zhang, C. M. and Corkum, P. B.},
   title = {Perturbing laser field dependent high harmonic phase modulations},
   journal = {J. Phys. B: At. Mol. Opt. Phys.},
   volume = {51},
   number = {12},
   ISSN = {0953-4075},
   DOI = {ARTN 125601
10.1088/1361-6455/aac18b},
   url = {<Go to ISI>://WOS:000432988700001
https://iopscience.iop.org/article/10.1088/1361-6455/aac18b},
   year = {2018},
   type = {Journal Article}
}

@article{Mairesse2005,
   author = {Mairesse, Y. and Quere, F.},
   title = {Frequency-resolved optical gating for complete reconstruction of attosecond bursts},
   journal = {Phys. Rev. A},
   volume = {71},
   number = {1},
   ISSN = {1050-2947},
   DOI = {ARTN 011401
10.1103/PhysRevA.71.011401},
   url = {<Go to ISI>://WOS:000227283300008
https://journals.aps.org/pra/pdf/10.1103/PhysRevA.71.011401},
   year = {2005},
   type = {Journal Article}
}

@article{Meng2026,
   author = {Meng, Lihui and Xu, Lu and Zhu, Xusheng and He, Lixin and Nie, Zan and Lan, Pengfei and Lu, Peixiang},
   title = {6.2-GW tabletop attosecond light source},
   journal = {arXiv},
   url = {https://arxiv.org/abs/2604.20084},
   year = {2026},
   type = {Journal Article}
}

@article{Meng2023,
   author = {Meng, L. H. and Liang, S. Q. and He, L. X. and Hu, J. C. and Sun, S. Q. and Lan, P. F. and Lu, P. X.},
   title = {Deep learning for isolated attosecond pulse reconstruction with the all-optical method},
   journal = {Journal of the Optical Society of America B-Optical Physics},
   volume = {40},
   number = {10},
   pages = {2536-2545},
   ISSN = {0740-3224},
   DOI = {10.1364/Josab.489019},
   url = {<Go to ISI>://WOS:001089172700004},
   year = {2023},
   type = {Journal Article}
}

@article{Nishimiya2025_arXiv,
   author = {Nishimiya, Kaito and Rajpoot, Rambabu and Takahashi, Eiji J},
   title = {Towards intense single-digit attosecond pulses with a 100-mJ-class mid-infrared sub-cycle laser},
   journal = {arXiv},
   url = {https://arxiv.org/abs/2510.27096},
   year = {2025},
   type = {Journal Article}
}

@article{Gagnon2008,
   author = {Gagnon, J. and Goulielmakis, E. and Yakovlev, V. S.},
   title = {The accurate FROG characterization of attosecond pulses from streaking measurements},
   journal = {Appl Phys B-Lasers O},
   volume = {92},
   number = {1},
   pages = {25-32},
   ISSN = {0946-2171},
   DOI = {10.1007/s00340-008-3063-x},
   url = {<Go to ISI>://WOS:000256908600006
https://link.springer.com/content/pdf/10.1007/s00340-008-3063-x.pdf},
   year = {2008},
   type = {Journal Article}
}

@article{Ivanov1996,
   author = {Ivanov, M. Y. and Brabec, T. and Burnett, N.},
   title = {Coulomb corrections and polarization effects in high-intensity high-harmonic emission},
   journal = {Phys. Rev. A},
   volume = {54},
   number = {1},
   pages = {742-745},
   ISSN = {2469-9926},
   DOI = {DOI 10.1103/PhysRevA.54.742},
   url = {<Go to ISI>://WOS:A1996UX03400087
https://journals.aps.org/pra/pdf/10.1103/PhysRevA.54.742},
   year = {1996},
   type = {Journal Article}
}

@article{Ammosov1986,
   author = {Ammosov, M. V. and Delone, N. B. and Krainov, V. P.},
   title = {Tunnel Ionization of Complex Atoms and Atomic Ions in a Varying Electromagnetic-Field},
   journal = {Zhurnal Eksperimentalnoi I Teoreticheskoi Fiziki},
   volume = {91},
   number = {6},
   pages = {2008-2013},
   ISSN = {0044-4510},
   DOI = {10.1117/12.938695},
   url = {<Go to ISI>://WOS:A1986F487200005},
   year = {1986},
   type = {Journal Article}
}

@article{Nishimiya2025,
   author = {Nishimiya, Kaito and Takahashi, Eiji J.},
   title = {Advances in dual-chirped optical parametric amplification},
   journal = {Advances in Physics: X},
   volume = {10},
   number = {1},
   pages = {2528351},
   ISSN = {null},
   DOI = {10.1080/23746149.2025.2528351},
   url = {https://doi.org/10.1080/23746149.2025.2528351},
   year = {2025},
   type = {Journal Article}
}

@article{DeLong1994,
   author = {DeLong, K. W. and Trebino, Rick and Kane, Daniel J.},
   title = {Comparison of ultrashort-pulse frequency-resolved-optical-gating traces for three common beam geometries},
   journal = {J. Opt. Soc. Am. B},
   volume = {11},
   number = {9},
   ISSN = {0740-3224
1520-8540},
   DOI = {10.1364/josab.11.001595},
   year = {1994},
   type = {Journal Article}
}

@article{Kane1999,
   author = {Kane, D. J.},
   title = {Recent progress toward real-time measurement of ultrashort laser pulses},
   journal = {IEEE J. Quantum Electron.},
   volume = {35},
   number = {4},
   pages = {421-431},
   ISSN = {0018-9197},
   DOI = {Doi 10.1109/3.753647},
   url = {<Go to ISI>://WOS:000079423800004
https://ieeexplore.ieee.org/stampPDF/getPDF.jsp?tp=&arnumber=753647&ref=},
   year = {1999},
   type = {Journal Article}
}

@book{Trebino2000,
   author = {Trebino, R.},
   title = {Frequency-resolved optical gating : the measurement of ultrashort laser pulses},
   publisher = {Kluwer Academic},
   address = {Boston},
   pages = {xvii, 425 p.},
   ISBN = {1402070667
9781402070662},
   url = {Contributor biographical information http://www.loc.gov/catdir/enhancements/fy0820/2002021855-b.html
Publisher description http://www.loc.gov/catdir/enhancements/fy0820/2002021855-d.html
Table of contents only http://www.loc.gov/catdir/enhancements/fy0820/2002021855-t.html},
   year = {2000},
   type = {Book}
}

@book{Hansen2010,
   author = {Hansen, Per Christian},
   title = {Discrete inverse problems : insight and algorithms},
   publisher = {Society for Industrial and Applied Mathematics},
   address = {Philadelphia},
   series = {Fundamentals of algorithms},
   pages = {xii, 213 p.},
   ISBN = {9780898716962
0898716969},
   url = {Table of contents only http://www.loc.gov/catdir/enhancements/fy1006/2009047057-t.html
Publisher description http://www.loc.gov/catdir/enhancements/fy1006/2009047057-d.html
Contributor biographical information http://www.loc.gov/catdir/enhancements/fy1006/2009047057-b.html
http://bvbr.bib-bvb.de:8991/F?func=service&doc_library=BVB01&doc_number=018939397&line_number=0001&func_code=DB_RECORDS&service_type=MEDIA Inhaltsverzeichnis},
   year = {2010},
   type = {Book}
}

@article{Trebino1997,
   author = {Trebino, R. and DeLong, K. W. and Fittinghoff, D. N. and Sweetser, J. N. and Krumbugel, M. A. and Richman, B. A. and Kane, D. J.},
   title = {Measuring ultrashort laser pulses in the time-frequency domain using frequency-resolved optical gating},
   journal = {Rev. Sci. Instrum.},
   volume = {68},
   number = {9},
   pages = {3277-3295},
   ISSN = {0034-6748},
   DOI = {Doi 10.1063/1.1148286},
   url = {<Go to ISI>://WOS:A1997XX25700001
https://watermark.silverchair.com/3277_1_online.pdf?token=AQECAHi208BE49Ooan9kkhW_Ercy7Dm3ZL_9Cf3qfKAc485ysgAABW0wggVpBgkqhkiG9w0BBwagggVaMIIFVgIBADCCBU8GCSqGSIb3DQEHATAeBglghkgBZQMEAS4wEQQMnIKZkrUjLgCFsNuqAgEQgIIFIHoWjNpxN1gd2HfV9BEKzeIUcR9UJ1EpkgZjpPkXUlgqvULr5tdgDjZVnZtZNUgcsp-n9mZc98W2j-mKkynKP-H-3ZSC8SPyPzQtkoBxUXYtj9YlVtX4T67frSa-nklujOb8SDqnEuEwVnWm1zPUrerYOMmXxEbrP-madBWz15KwPkkGc9TmyFg1bTz0jqrsb3ZoR6oKCOtUne_Nle4vJfK9oBlNnzYn2qQM9VHcYiag5bpvIf5aND4a5f2ui-rrcknMKlJ9egDUc2RmH7ZRADu9RJK5TE3YMAyQqRMjVQWxJHhzHYYyZmUTEPTRpOE1yIBGuTa89-Mp-hOMDGc9rMyBeTSqRgc9jnPNfCMpqoSc-9T70A7x84U7H1I68UCU6N3Jbw8G21bCH7Se8UPpoxmwh1xs9uN1LtfSJOV1nKbvKxvOSe8bcHHtKOqh0MQpP8K_ieo7gnfFRXykJnmLGsy5mnLsM7m7TBiQa-Ecf3_Gw1vvC5XoafOJ5Y-cOmBXsZGk78bp8xqtQNVNxvRuko53yF3nVbLXLfAq-vWm_htyQY2wVPVCGrLFmUpBP7LmRKVDpyxZy7_okZoR-CCT29qOBAtNwLzO0dL9puZHuKymFivtHaHecrf2QeFTPF_rPfdYgjSCyNb8LqN78J3aJmWW_ntPvoUBswRmO2HPffvmqKbB3Hk6wJ6OdsI8aocxj1qw-2iIDL2YPi3_z8yTi56rOWjOyRIowaiPSSPjgdpYaVgxLRpik33kpKmExCO9yZiMM3Jh-5idDeNquTb7hzrAtC8fx6dpgpxvRlqyHOoA15YMOfKjHLAbACznQQ4t-6Ze55iJ6Zsuih8Am9_Qs8w7aTseskgYksiBUQN8uvlzRUiZ1QlAT69ZYlPsSpGZzAOTzjRoPsp-r-DXT0VFgroNjNqhIG2YULwk2vDZsHr7z_UFy9JJzLaSeFQk8N6ymChYo72fKIO8BdLjTuV9rJ35eAxka7xgrQcfk3A0yakcig33tOv1T0Om5AaC1sYX0aNguidAkitnwIhocMw9zf3Q22sLXQNotQ0AxCW8vFxRL1U39pPOwRFWSaQFfMlIRJIArGEUWl7ItQLkOjOYkPKJApuf0NzYT600dHGiQSDB9mmtyW1aqnz390ZIjloi_QvXlEfSzSB5LvAhV_0axF8HNhs3HhCF1FZYa5j64TQx6vXRpIQiFJ2BlBeZGZ20cGz-BUa_FZyeY77TIR0T_x-KlBnKAGX1ToyCx1xjOS3iTqOvBFprC9zA4kTUmnDsbj9rGeF20WR249Xr5iiDXYSnq7bwRsJlohQYS__47WEytVaglR0dTMwkyomAfqBM4EkWavayalUuG-I1oRWqEPeJz1U6UFIrR2u4P4lS5-K7_f1APbqbrJpYbShyFInDZkx7eEd_jV-cZTe_XxWtLKwj6LSSN5zaaB3xlbs6JLNbDN-fTeIlQP0GqqIrac_dU0pp7dRMIU9gOoZvNUSt850spoVvd44txoE4I1Co_Gtf1JosmjGvX5YU2y-fxn8-GfAgmAbjSojTftOsMl4Kimi36oB9yd1d6gtBZbI45Hv015wd_DIib6p6GSd6ACO4PCNCHeZ3OwXgNaHgiD6n60pWYAp_-I68rKA8QheQo8Gg0S8N-2REWqYGGpaBAGmQqRQdgcMo4XtjgPUw18efA3YEG6ioaA6R1qXB75NDCHariRN_zb-ycyezondw57d5iSURBQFBdM0BuzFQE1qEHjk},
   year = {1997},
   type = {Journal Article}
}

@article{Wei2024,
   author = {Wei, Z. J. and Yan, M. D. and Xia, F. and Men, T. and Tang, W. Q. and Yan, W. W. and Liu, S. Y. and Li, Z. Y.},
   title = {Relative strengths of sum and difference frequency generation in perturbative high harmonic wave mixing},
   journal = {Journal of the Optical Society of America B-Optical Physics},
   volume = {41},
   number = {6},
   pages = {1347-1354},
   ISSN = {0740-3224},
   DOI = {10.1364/Josab.525386},
   url = {<Go to ISI>://WOS:001241017800009},
   year = {2024},
   type = {Journal Article}
}

@article{Yan2025,
   author = {Yan, M. D. and Wei, Z. J. and Gao, X. Z. and Lu, J. H. and Li, Z. Y.},
   title = {Amplitude modulation effect on in-situ temporal characterization of high harmonic attosecond pulses},
   journal = {Opt. Express},
   volume = {33},
   number = {7},
   pages = {16620-16630},
   ISSN = {1094-4087},
   DOI = {10.1364/Oe.559512},
   url = {<Go to ISI>://WOS:001467897000004},
   year = {2025},
   type = {Journal Article}
}

@article{Yang2020,
   author = {Yang, Z. and Cao, W. and Chen, X. and Zhang, J. and Mo, Y. L. and Xu, H. Y. and Mi, K. and Zhang, Q. B. and Lan, P. F. and Lu, P. X.},
   title = {All-optical frequency-resolved optical gating for isolated attosecond pulse reconstruction},
   journal = {Opt. Lett.},
   volume = {45},
   number = {2},
   pages = {567-570},
   ISSN = {0146-9592},
   DOI = {10.1364/Ol.381188},
   url = {<Go to ISI>://WOS:000510865100078},
   year = {2020},
   type = {Journal Article}
}

\end{document}